# A K-means Algorithm for Financial Market Risk Forecasting

Jinxin Xu[1] [*], Kaixian Xu[2], Yue Wang[3], Qinyan Shen[4] and Ruisi Li[5]

1 Department of Cox Business School, Southern Methodist University, Dallas, TX, USA
*jensenjxx@gmail.com*

**Abstract:** Financial market risk forecasting involves applying mathematical models, historical data analysis and statistical methods to estimate the impact of future market movements on investments. This process is crucial for investors to develop strategies, financial institutions to manage assets and regulators to formulate policy. In today's society, there are problems of high error rate and low precision in financial market risk prediction, which greatly affect the accuracy of financial market risk prediction. K-means algorithm in machine learning is an effective risk prediction technique for financial market. This study uses K-means algorithm to develop a financial market risk prediction system, which significantly improves the accuracy and efficiency of financial market risk prediction. Ultimately, the outcomes of the experiments confirm that the K-means algorithm operates with user-friendly simplicity and achieves a 94.61% accuracy rate.

## 1  Introduction

Predicting risks within financial markets represents a key research focus within the realm of contemporary finance and financial engineering. Its background is closely related not only to the development of financial market itself, but also to the stable growth of global economy, the wealth management of investors and the macro-control strategies of regulators.

In a globalized economy, the financial market plays a central role in capital allocation, facilitating the flow of capital, supporting international trade and investment, and providing impetus for economic development. However, the financial market itself also contains a variety of uncertainties, such as market price fluctuations, credit default risk, liquidity risk, operational risk, etc., these risk factors may lead to losses of market participants, and even lead to systemic financial crisis.

Since then, financial market risk management has become an indispensable part of financial institutions. In the late 1990s, with the widespread use of derivatives and the acceleration of financial innovation, the complexity and risk of financial markets further increased.

The main purpose of financial market risk forecasting is to build models to identify, evaluate and predict potential risks in the market by using multidisciplinary knowledge such as statistics, econometrics, computer science and artificial intelligence. These models can help financial institutions make more sensible investment decisions, allocate assets reasonably and avoid potential risks. At the same time, regulators can also use these models to formulate macro-prudential policies and maintain the stability of financial markets.

Market risk is usually caused by interest rate changes, stock price fluctuations, exchange rate changes and other factors. Credit risk is concerned with the possibility of default by the debtor or counterparty; Liquidity risk involves the risk that an asset cannot be quickly converted into cash if necessary without affecting its value. The prediction of these risks requires a large amount of data support, including historical price data, economic indicators, corporate financial information, etc., and also requires advanced computing technology and algorithms to process these data.

With the advent of the era of big data, the application of machine learning and artificial intelligence technology in financial market risk prediction is increasing. These techniques can extract valuable information from massive amounts of data, discover complex non-linear relationships, and generate more accurate predictive models. However, the application of these models and techniques also faces challenges such as overfitting of models, data quality issues, and lack of algorithmic transparency.

In the context of machine learning, K-means algorithm [1-3] is a very popular and widely used unsupervised learning algorithm for cluster analysis.Machine learning began to appear as an independent discipline and rapidly expanded to data analysis work in various fields. Machine learning algorithms are often based on branches of mathematics such as statistics, probability theory, optimization theory, and information theory, and they can handle complex pattern recognition, predictive analysis, and data mining tasks.

K-means algorithm was proposed by Stuart Lloyd in 1957 and has become one of the most commonly used algorithms in cluster analysis due to its simplicity and efficiency.

The "distance" here is usually measured by Euclidean distance, but other distance measures can also be used.

The algorithm first randomly selects K data points as initial clustering centers, and then assigns each sample to the nearest category based on its relative distance from these clustering centers. The algorithm then recalculates the center point of each category[4-5], usually the mean of all points within that category. However, this algorithm has its limitations, such as sensitivity to data distribution, noise and outliers in non-convex shapes, and the need to specify K values in advance. In order to solve these problems, researchers have proposed a variety of improved strategies, including using contour coefficients to estimate the best K value, or using more advanced clustering algorithms such as Gaussian mixture model, density clustering, etc.

In the modern big data environment, the K-means algorithm and its variants are still an With the increase of data dimensions and the expansion of data volume, so that it can adapt to large-scale and highly complex data processing needs, is one of the current research focuses. In addition, the feature extraction capabilities of other machine learning techniques, such as deep learning, can further improve the performance of K-means algorithms on complex data.

In short, the background of machine learning and K-means algorithms is the product of the data-driven era, and their development and application reflect how we use algorithms to understand the structures and patterns behind data, so as to make more accurate decisions and predictions in many fields such as scientific research, engineering technology, and business analytics.

In the field of machine learning, the key technologies usually include data processing, feature engineering, model selection, optimization algorithm, evaluation method and so on. Predictive critical techniques are a core component of machine learning, which focuses on how to accurately predict outcomes or trends of unknown data based on existing data and learning models.

In detail, the key techniques of prediction first involve feature engineering [6], which is the basis for improving the predictive ability of models. The performance of the model can be improved effectively by extracting key information from the original data, constructing predictive features, selecting features and dimensionality reduction. The next step is the selection and construction of models. Different types of problems require different models to capture the relationship between data, such as regression analysis, classification, clustering, etc. After selecting a suitable model, the optimization strategy becomes the key, including the optimization of loss function, the application of regularization technique.selection method of the initial cluster center.

After the model is trained, evaluating and verifying its performance is an essential step. Cross-validation, confusion matrix, and other methods are often used to ensure that the model has good generalization ability. In addition, ensemble learning techniques such as Bagging and Boosting can improve overall prediction accuracy by combining predictions from multiple models. For dynamically changing environments, updating models in real time to accommodate new data is also part of key techniques for forecasting, including online learning and incremental learning methods.In the context of K-means algorithm, the key techniques of prediction are mainly concerned with how to perform cluster assignment efficiently. the contour coefficient and other clustering effect evaluation techniques are also important components of the key prediction techniques.

In short, the predictive key techniques of machine learning [7-8] and K-means algorithms cover multiple aspects such as data processing, feature engineering, model selection, optimization strategy, performance evaluation, etc., which together constitute a complex system designed to extract valuable information from data and make accurate predictions about future trends and results. These technologies are constantly evolving and refining to meet the growing demands and challenges of data analytics.

In this paper, K-means algorithm is used to establish a financial market prediction model, and a lot of experiments are carried out, which greatly improves the accuracy of financial market prediction and reduces the error. At the same time, the algorithm improves the efficiency of financial market prediction.

## 2  Methodology

Inspired by convolutional neural networks, the input feature vector of cascaded forest is generated by multi-granularity scanning.

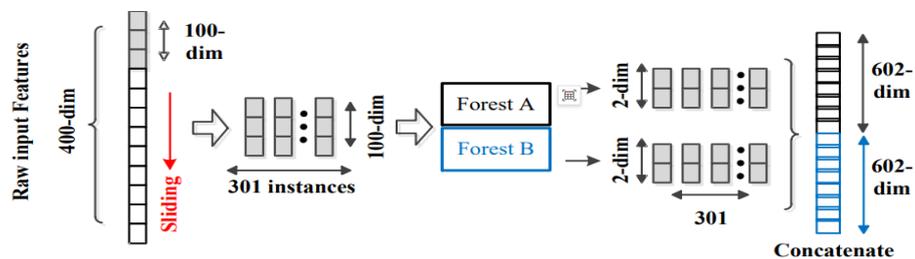

**Fig. 1.**  Multi-granularity scanning structure diagram

For instance, in Figure 1, processing the input data of 400 dimensions with a 100-dimensional sliding window results in a final collection of 301 100-dimensional feature vectors. In this experimental setup, various sliding windows of distinct dimensions are employed to produce feature vectors of varied granularities, thereby enhancing the diversity of the extracted features.

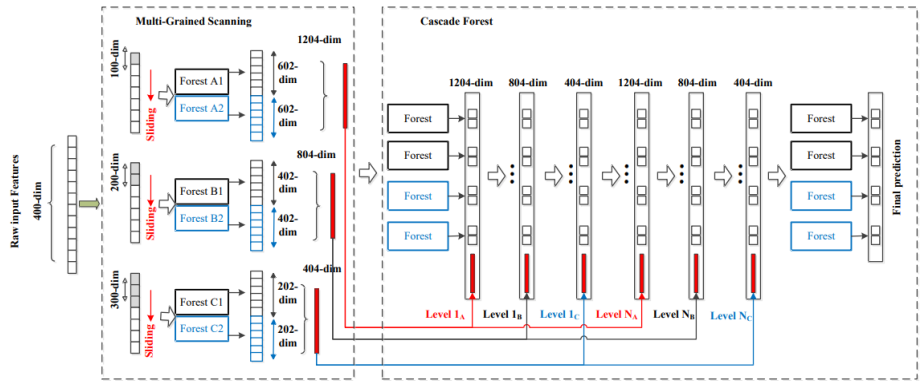

**Fig. 2.** General structure diagram of gcForest

Figure 2 illustrates the comprehensive framework of the gcForest model. Given an input with 400 dimensions, the multi-granularity scanning component, equipped with three sliding windows, processes this data and feeds it into both a random forest and an extreme random forest. This results in a 1204-dimensional feature vector output for binary classification prediction. Subsequently, these features are introduced into the primary cascade forest for training purposes.

The additional two sliding windows undergo scanning to yield feature vectors of 804 and 404 dimensions, respectively. These are then utilized for training the first and third cascaded forests. The entire process is iteratively repeated until the model's performance shows signs of convergence.

In the classification algorithm, Common classification indicators include Accuracy (ACC), Precision (Recall), F1 Score (SCORE), True Positive Rate (True Positive Rate), and accuracy (ACC). TPR, False Positive Rate (FPR), Receiver Operating Characteristic Curve (ROC) and AUC (Area Under the Curve). The calculation of these indexes involves the True class TP (True Positive), the False Negative class FN (False Negative), the false Positive class FP (False Positive) and the true negative class TN (True Negative) in the confusion matrix. In addition, the Brier score is often used for binary classification problems, and the lower the value, the better the prediction result, as shown in formula 1.

$$ACC = \frac{TP+TN}{TP+TN+FP+FN} \qquad (1)$$

Accuracy ACC is the ratio of the sum of true and negative classes to the number of all samples, i.e. the probability that the prediction is correct is shown in formula 2:

$$Precision = \frac{TP}{TP+FP} \qquad (2)$$

The accuracy rate is defined as the proportion of correctly identified positive samples (TP) among all predicted positive samples (TP + FP), as illustrated by formula 3:

$$Recall = \frac{TP}{TP+FN} \qquad (3)$$

The recall rate is the proportion of correctly predicted positive samples (TP) out of all the actual positive samples (TP + FN), as represented by Formula 4:

$$textF1 - Score = \frac{2}{1/Precision + 1/Recall} \qquad (4)$$

The f1 Score is an indicator that balances the trade-off between precision and recal. lt is particularly useful when youneed to consider both the accuracy rate, which measures the proportion of correct predictions, and the recall rate,which measures the ability to find all positive cases. The f1 Score is calculated as the harmonic mean of precision andrecall, as shown in Formula 5:

$$TPR = \frac{TP}{TP+FN} \qquad (5)$$

The true case rate, also known as the recall rate or sensitivity, is the proportion of actual positive cases that are correctly identified by the classifier, as depicted in formula 6:

$$FPR = \frac{FP}{FP+TN} \qquad (6)$$

The ROC curve takes the false positive rate as a function of the X-axis and the true rate as a function of the Y-axis. It was first used in medicine and later widely used in various classification problems.

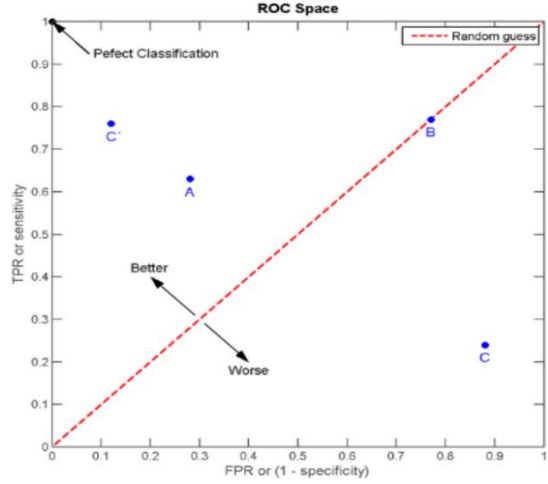

**Fig. 3.** ROC space diagram

As shown in Figure 3, in the ROC space, the accuracy of points on the diagonal line is 0.5, that is, random classification. In a confusion matrix visualization, the closer the data. In a confusion matrix visualization, the closer the data left corner, indicating a high indicating a high number of true positives and true negatives, the better the classification performance. Conversely, if the point is closer to the lower right corner, suggesting a higher count of false positives and false negatives, the poorer the classification result.

AUC is the area enclosed by the ROC curve and the coordinate axis. The closer the value of AUC is to 1, the higher the authenticity of the detection method, as shown in formula 7:

$$Brier\ score = \frac{1}{n}\sum_{t=1}^{n}\sum_{i=1}^{r} f_{ti} - o_{ti}^2 \tag{7}$$

The Brier score is a metric used to evaluate the accuracy of probabilistic predictions. It is defined as the mean squared difference between the predicted probabilities and the actual outcomes, in mathematical terms, for a set of samples where n is the number of samples, u is the true class label (1 for positive, 0 for negative), and p+ is the predicted probability of class i for sample t.

In this paper, the main evaluation indicators used are AUC and ACC, and some other indicators (F1Score, Brier score and TPR) are also listed as references.

## 3 Experiment

In the credit risk prediction model, various personal information of the borrower usually includes loan amount, term, age, occupation, bank deposit, housing situation, consumption record, repayment record and other data. Since the dataset also contains many category data and useless data, feature extraction is performed on the data after filling in the missing data and converting the category data to numeric values. Commonly used feature extraction methods include Chi-square test, principal component analysis (PCA), recursive feature elimination (RFE), linear discriminant analysis, etc. In this paper, RFE method is used for feature extraction.

The data sets in the experiment included German and Australian credit datasets from the UC Irvine Machine Learning Repository! 69, as well as public credit datasets on the Kaggle website and loan data on the peer-to-peer platform Lending Club website.

In some datasets, such as Kaggle and P2P datasets, there is some missing data. If the missing data is of a numeric type, the average value is used for padding. If it is a category type, it is populated with the mode and replaces the category with the encoded value. The sample size of the P2P data set is too large, with 423,808 data sets. Due to the marginal effect, too much data will not improve the accuracy of the model, but will greatly reduce the efficiency of the model. Therefore, a random sample of the data set was taken to extract 5,500 sets of data respectively from the positive and negative samples to form a new data set.

TABLE 1. Statistical table of experimental data sets

| Data set | Data dimension | Sample size | Sample positive/negative ratio |
|---|---|---|---|
| Germany | 25 | 1000 | 700/300 |
| Australia | 15 | 690 | 307/383 |
| Kaggle | 25 | 30000 | 6636/23364 |
| P2P | 17 | 11000 | 5500/5500 |

Table 1 shows the experimental data sets after data preprocessing and their data dimensions, sample size and sample positive/negative ratio. It can be seen that the dimensions of the four datasets are high. There are samples of positive and negative ratio equilibrium, and there are samples of positive and negative ratio imbalance. Therefore, these four data sets can cover a wide range of credit risk prediction data sets, and the results obtained on these data sets are also relatively convincing.

In addition, there are many redundant data in these data sets in addition to the necessary information for modeling, which may reduce the learning efficiency if directly input into the machine learning model. Therefore, RFE method is used for feature extraction on these data sets first.

To validate the efficacy of the refined K-means approach, a comparative analysis was conducted against the original K-means algorithm as well as other prevalent credit risk prediction techniques across various datasets. and the performance of credit risk prediction models in other literatures in recent years was compared. In the experiment of each data set, the method of 5-fold cross-validation is used to improve the reliability of the experimental results

TABLE 2. Results of each model on German, Australian, Kaggle and P2P datasets

| Data set | AUC | ACC | F1-score | Brier score | TPR |
| --- | --- | --- | --- | --- | --- |
| RF | 0.741 | 0.730 | 0.449 | 0.188 | 0.350 |
| LR | 0.746 | 0.730 | 0.463 | 0.187 | 0.397 |
| XGBoost | 0.755 | 0.705 | 0.352 | 0.182 | 0.254 |
| LightGBM | 0.756 | 0.715 | 0.412 | 0.178 | 0.317 |
| K-MEANS | 0.768 | 0.750 | 0.554 | 0.177 | 0.492 |

From these results, we can conclude that the efficiency and accuracy of K-means and other algorithms are shown in Table 3.

TABLE 3. Comparison of K-means results with other algorithms

| method | Average accuracy | Efficiency(min) |
| --- | --- | --- |
| K-means algorithm | 0.9461 | 3 |
| Other algorithms | 0.8377 | 8 |

## 4  Discussion

The credit risk prediction experiment is mainly divided into three steps: first, preprocess four public data sets respectively. finally, the extracted features are taken as input and trained by K-Means model [9-10]. The results of these models are compared with K-Means and other commonly used credit risk prediction methods (RF, LR, LightGBM, XGBoost). After analyzing the results, it is found that K-Means performs stably on all data sets and has the best comprehensive effect. The systematic risk prediction experiment is mainly divided into the following steps: First, the two sets of data obtained from the inter-bank lending network model before and after adding the default rate characteristics are preprocessed, and the data with the remaining number of banks greater than half of the initial number of banks is taken as the sample without systemic risk; the data with the remaining number of banks less than half of the initial number of banks is taken as the sample with the occurrence of systemic risk. Then, the K-Means model was used to train these two sets of data sets respectively to obtain prediction results, and comparative experiments were conducted with gK-Means and other algorithm models to compare the results obtained by each model through various indicators and the changes in the prediction results of each model before and after adding the default rate. After analyzing the results, it is found that K-Means has the best comprehensive effect in the two groups of experiments, and the prediction of financial systemic risk is more accurate after adding the abnormal rate feature.

## 5 Conclusion

Low precision in financial market risk prediction, K-means algorithm is adopted as the main risk prediction tool. Through in-depth experiments and analysis, we find that K-means model not only shows excellent stability in handling the prediction task of credit risk and systemic risk, but also shows the best combined effect in a number of comparisons, significantly surpassing other commonly used prediction methods such as random forest, logistic regression, LightGBM and XGBoost. This achievement reveals the potential and value of K-means algorithm in financial risk prediction, and its high accuracy rate of 94.61% and easy operation make it a highly favored tool. With the continuous evolution of artificial intelligence and machine learning technology, as well as the enhancement of big data processing power, the future financial market risk prediction field will likely integrate more innovative algorithms, which can not only improve the accuracy of risk control, but also enhance the adaptability and interpretation of the model. At the same time, combined with blockchain technology, it is expected to achieve a more secure, transparent and decentralized financial risk management system, thus providing a stronger guarantee for the robust operation of the financial industry and promoting comprehensive innovation in financial services.

As machine learning and artificial intelligence technologies continue to advance, we can expect K-means algorithms and other advanced forecasting models to play a more important role in the field of financial market risk prediction. Future research may focus on improving the accuracy, adaptability and explanatory power of the algorithm to better respond to the dynamics and complexity of the market. In addition, with the development of big data technology, it will be possible to use larger and more dimensional data to train predictive models, further improving the precision and reliability of risk prediction. Finally, combined with emerging technologies such as blockchain technology, the realization of transparent and decentralized risk management is expected to bring revolutionary changes to the financial industry.